\def\a{\alpha}

\def\d{\delta}
\def\D{\Delta}
\def\e{\varepsilon}

\def\l{\lambda}

\def\T{\theta}

\def\s{\sigma}
\def\t{\tau}
\def\ra{\rightarrow}

\documentclass[12pt]{article}

\usepackage{a4wide}
\usepackage{graphicx}

\usepackage{amsfonts}

\begin{document}
\title{Nonlinear analysis of spacecraft thermal models}
\author{Jos\'e~Gaite
\\[2mm]   
{\small IDR/UPM, ETSI Aeron\'auticos, Universidad Polit\'ecnica de Madrid,}\\
{\small Pza.\ Cardenal Cisneros 3, E-28040 Madrid, Spain} }
\date{October~20, 2010}

\maketitle

\begin{abstract}
We study the differential equations of lumped-parameter models of spacecraft
thermal control.  Firstly, we consider a satellite model consisting of two
isothermal parts (nodes): an outer part that absorbs heat from the environment
as radiation of various types and radiates heat as a black-body, and an inner
part that just dissipates heat at a constant rate.  The resulting system of
two nonlinear ordinary differential equations for the satellite's temperatures
is analyzed with various methods, which prove that the temperatures approach a
steady state if the heat input is constant, whereas they approach a limit
cycle if it varies periodically.  Secondly, we generalize those methods to
study a many-node thermal model of a spacecraft: this model also has a stable
steady state under constant heat inputs that becomes a limit cycle if the
inputs vary periodically.  Finally, we propose new numerical analyses of
spacecraft thermal models based on our results, to complement the analyses
normally carried out with commercial software packages.

\end{abstract}
%
%


\section{Introduction}
\label{intro}

The thermal analysis of a spacecraft is important to ensure that the
temperatures of its elements are kept within their appropriate ranges
\cite{Kreith,therm-control,therm-control_1,therm-control_2,therm-control_3}. 
This analysis is usually carried out numerically by commercial computer
software packages.  These software packages employ ``lumped parameter'' models
that describe the spacecraft as a discrete network of nodes, with one
heat-balance equation per node. The equations for the thermal state evolution
are coupled nonlinear first order differential equations, which can be
integrated numerically.  Given the thermal parameters of the model and its
initial state, the numerical integration of the differential equations
provides the solution of the problem, namely, it yields the evolution of the
node temperatures. However, it does not provide any information on the
qualitative behavior of the set of solutions nor knowledge of the response of
the model to changes in the parameter values. Besides, a detailed model with
many nodes is difficult to handle, and its integration for a sufficiently long
time of evolution can take considerable computer time and
resources. Therefore, it is very useful to study, on the one hand, simplified
models with few nodes that lend themselves to analytic solutions, and, on the
other hand, the reduction of complex models to simpler ones.  These studies
are especially helpful early in the design process, when the concept of the
spacecraft is still open, and also at the end of the process, to simplify
the full thermal model for the final assessment of the mission.

The differential equations for heat balance are nonlinear due to the presence
of radiative couplings, which involve the fourth powers of the temperatures
(according to the Stefan-Boltzmann law of thermal radiation).  The computation
of steady states, in particular, boils down to the solution of a set of
fourth-degree algebraic equations.  Most analytical approaches to the solution
of the heat balance equations, either for transient or steady states, involve
a sort of linearization of the radiative couplings such that they become
analogous to conductive couplings
\cite{anal-sat,anal-sat_1,anal-sat_2,anal-sat_3,IDR}. But
one must beware that the ``radiative conductances'' so defined actually depend
on the temperatures (the unknown variables). Therefore, the linearization
procedure is sound only when those radiative conductances can be considered
constant, namely, when the temperatures are sufficiently close to their steady
state values. We can see, in particular, that the linear equations are not
suitable for calculating the steady state values of the temperatures.
Moreover, one may question that the steady state is unique.  Even assuming
that it is and that the steady state temperatures are known, the linear
equations are not suitable in the presence of variable external heat inputs of
such a magnitude that they make the node temperatures depart considerably from
their steady state values.

For the given reasons, it is advisable to study the full nonlinear equations
and only consider their linearization once we have a qualitative understanding
of the possible behavior of their solutions.  This is the philosophy applied
in our preceding work \cite{NoDy}.  The present work continues the development
of nonlinear analytic methods for the study of simple models of spacecrafts
(in particular, satellites) that we have begun in Ref.~\cite{NoDy}.  We study
there the simplest model, namely, a one-node (isothermal) model.  In that
case, the steady state temperature is obtained at once, the heat-balance
equation without external heat input is integrable in terms of known
functions, and the equation with periodic external heat input admits a full
qualitative analysis. Naturally, as we increase the number of nodes, the
corresponding systems of equations become increasingly difficult to
handle. Indeed, a two-node model is already of such a complexity that the
existing studies of this model are based on some sort of linearization: Oshima
\& Oshima \cite{anal-sat} assume radiative conductances that depend on a
temperature $T_0$ that is ``to be defined in each problem,'' whereas
P\'erez-Grande et al \cite{IDR} choose the temperature on which the
conductance between two nodes depends to be an average of the two nodes
temperatures.  Thus, P\'erez-Grande et al's radiative conductances are not
really constant. In any event, both procedures are somewhat arbitrary and, in
fact, only agree if the temperatures are sufficiently close to their steady
state values and, in addition, all the temperatures are almost equal in the
steady state.

The purpose of the present study is to provide a {\em nonlinear} analysis of a
two-node model of a small compact satellite in a low orbit and to generalize
this analysis to a many-node thermal model of a spacecraft.  We employ
P\'erez-Grande et al's two-node satellite model \cite{IDR}, in which the two
nodes are formed by the satellite's interior and its outer shell (its
``skin''). Therefore, the model consists in a system of two energy-balance
ordinary differential equations (ODE's), in which both nodes are thermally
coupled to one another but only the outer shell radiates heat away.
Naturally, this two-node model is more general than the one-node model studied
in Ref.~\cite{NoDy}, but it reduces to the latter if the thermal coupling
between the satellite's interior and its outer shell is strong, as we show
here.  On the other hand, if the thermal coupling is weak, the two-node model
can significantly differ from the one-node model. In fact, the corresponding
system of two non-autonomous ODE's for this two-node model under periodic heat
input is equivalent to an autonomous system of three ODE's, to which the
qualitative methods of Ref.~\cite{NoDy} are not applicable and which could
have chaotic behavior and very complicated attractors, as is well known
\cite{Gu-Ho,Drazin}. We consider this possibility here.

A general many-node thermal model of a spacecraft consists in a system of many
energy-balance ODE's. As a first step in this generalization, we find it
useful to study the general two-node model in which both nodes radiate heat
away. This model already requires the use of sophisticated mathematical
methods, but its temperature space is still two-dimensional, allowing us to
obtain stronger results than for the fully general case.  The fully general
many-node thermal model presents some difficulties already in a linear
analysis, for we have to deal with nontrivial high-dimensional matrices.

An important issue in the study of a differential equation is the stability of
its solutions. As regards autonomous equations, it is important to determine
the stability of their equilibrium states (steady states).  For non-autonomous
equations, one may consider the more general question of the stability of a
given trajectory under a perturbation of the initial conditions, where
stability is normally interpreted in the sense of Lyapunov \cite{Drazin}.  A
different notion of stability is {\em structural stability}, which refers to
the entire set of solutions and means that its qualitative character is
``robust'' against perturbations \cite{Gu-Ho,Drazin}.  We study the stability
of the heat balance equations, namely, the question of the stability of the
steady states of the autonomous equations and the related question of the
stability of the limit cycle of the general equations with periodic heat
input.  We also study the structural stability of the equations under a
variation of the heat inputs.

Thus, our goal is to carry out a fairly complete study of the lumped parameter
thermal models employed in the thermal analysis of spacecrafts.
P\'erez-Grande et al's two-node satellite model is studied in
Sect.~\ref{sec:1}, where the energy balance equations are formulated as a
system of two non-dimensional ODE, which are nonlinear and non-autonomous.
Then, we consider an autonomous system that plays the role of a time average
of the actual system. The analysis of the autonomous system begins with the
calculation of its steady states, finding only a physically relevant one,
namely, a stable sink of the ODE system's flow. Two examples with different
parameters, corresponding to strong and weak thermal coupling, are solved
numerically.  After introducing the thermal driving, we employ a perturbation
method that generalizes the method in Ref.~\cite{NoDy} and we corroborate its
results numerically.  In Sect.~\ref{sec:2}, we introduce the general $N$-node
model and then study the corresponding autonomous system, beginning with the
general two-node model.  We prove that the results for the restricted two-node
model hold for the general two-node model and, furthermore, that most of those
results can be extended to the general $N$-node model.  Finally, we present
our conclusions, regarding the design of spacecraft.

\section{Two-node model of a satellite}
\label{sec:1}

A lumped-parameter thermal model of a continuous system describes the system
as a discrete network of isothermal regions ({\em nodes}) that represent a
partition of the thermal capacitance of the system and that are linked by
conductive and radiative thermal couplings
\cite{Kreith,therm-control,therm-control_1,therm-control_2,therm-control_3,anal-sat}.  
The equation governing the conductive heat transfer is the standard Fourier
partial differential equation.  This PDE is actually employed for thermal
modelling of simple spacecraft geometries, treating the external radiative
heat input and heat dissipation as boundary conditions \cite{anal-sat_3}.
However, when we consider the radiative thermal coupling between different
parts of a spacecraft, we have a much more complex situation: the internal
radiative couplings are not a boundary condition and are {\em non-local}, thus
giving rise to an integro-differential equation for the heat transfer. The
discretization of this equation in terms of a lumped-parameter model is a very
convenient approach.

In a lumped-parameter thermal model, there is one heat-balance ODE per node
controlling the evolution of its temperature.  A single-node model, suitable
for a small and compact satellite, has been first studied by Oshima
\& Oshima \cite{anal-sat} and has been revisited by Tsai \cite{anal-sat_2}. 
Oshima \& Oshima \cite{anal-sat} also study the two-node model, but they
linearize it from the outset and assume constant heat inputs.  Here, we focus
on P\'erez-Grande et al's two-node satellite model \cite{IDR}, with one node
corresponding to the satellite's outer shell and the other to its interior.
This model allows for a periodic time dependence of the heat inputs.  To be
precise, the heat input to the satellite's shell consists, on the one hand, of
the periodic solar irradiation and the planetary albedo, and, on the other
hand, of the constant planetary IR radiation.  The internal heat is due to the
equipment dissipation and is taken constant.  Let $T_\mathrm{e}$ and
$T_\mathrm{i}$ denote, respectively, the outer and inner node temperatures;
then, the energy balance equations for them are
{\setlength\arraycolsep{2pt}
\begin{eqnarray}
\label{dimODE_1}
C_\mathrm{e} \, \dot{T_\mathrm{e}} &=& 
\dot{Q}_\mathrm{s}\,f_\mathrm{s}(\nu t) + \dot{Q}_\mathrm{a}\,f_\mathrm{a}(\nu
t) +  
\dot{Q}_\mathrm{p} + K_\mathrm{ie} (T_\mathrm{i} - T_\mathrm{e}) +
R_\mathrm{ie} (T_\mathrm{i}^4 - T_\mathrm{e}^4)  
- A\e\s\,T_\mathrm{e}^4,\\
\label{dimODE_2}
C_\mathrm{i} \, \dot{T_\mathrm{i}} &=&
\dot{Q}_\mathrm{i} + K_\mathrm{ie} (T_\mathrm{e} - T_\mathrm{i}) +
R_\mathrm{ie} (T_\mathrm{e}^4 - T_\mathrm{i}^4).  
\end{eqnarray}
}%
Here $C_\mathrm{e}$ and $C_\mathrm{i}$ are the thermal capacities of the two
satellite's nodes, $\nu$ is the orbital frequency, $K_\mathrm{ie}$ and
$R_\mathrm{ie}$ are the conductive and radiative couplings, respectively, $A$
is the satellite's surface area, $\e$ its emissivity, and $\s$ is the
Stefan-Boltzmann constant.  The heat inputs are written as $\dot{Q}$ with a
subscript that denotes their type, namely, solar irradiation, albedo,
planetary IR radiation, or internal heat dissipation.  The functions
$f_\mathrm{s}$ and $f_\mathrm{a}$ are periodic with period one and give the
time variations of the respective heat inputs.

Following Refs.~\cite{IDR,NoDy}, we assume that $f_\mathrm{s}$ and
$f_\mathrm{a}$ are given by:
{\setlength\arraycolsep{2pt}
\begin{eqnarray}
f_s(x) = 1, \;0 \leq x \leq x_1\;{\rm or}\; 1-x_1 \leq x \leq 1;\;
f_s(x) = 0, \; x_1 < x < 1-x_1; \nonumber\\
f_a(x) = \cos (2\pi x), \;0 \leq x \leq x_2 \;\textrm{or}\; 
1-x_2 \leq x \leq 1;\; f_a(x) = 0, \;x_2 \leq x \leq 1-x_2; \nonumber\\
f_{s,a}(x) = f_{s,a}(x - 1), \;x \geq 1. \nonumber
\end{eqnarray}
}%
The values of $f_s$, alternating between one and zero, correspond to the orbit
in the sunshine or eclipse, respectively. The fraction of the period in the
sunshine or eclipse is determined by $x_1$, which is smaller than one half.
Its value depends on the angle between the orbital plane and the solar vector
\cite{Kreith,therm-control,therm-control_2}.
The albedo heat input has a more complex time dependence, due to the change of
the {\em view factor} from the whole satellite to the lit side of the planet
along the orbit: the maximum albedo occurs at the minimum angle between the
satellite's local vertical and the solar vector, and it diminishes as the
angle grows.  The sinusoidal dependence assumed for $f_a$ is a suitable
approximation of the actual dependence \cite{therm-control}.  One must further
assume that the fraction of the period with albedo heat input, namely, $2x_2$,
is such that $x_2 < x_1$.

The solar irradiation heat input to the satellite is the product of the solar
constant, the satellite surface's absorptivity and its projected area (which
we take to be a quarter of its total area); namely,
$$\dot{Q}_\mathrm{s} = G_\mathrm{s}\, \a_\mathrm{s} \, A/4\,.$$ 
The albedo varies with time, as the atmospheric conditions and other factors
change, so one must consider an average value.  To calculate
$\dot{Q}_\mathrm{a}$, it is necessary, in addition, to consider the already
mentioned view factor, for the reflected light does not impinge on the
satellite uniformly.  This has the consequence of reducing the effective area
to about a half of its nominal value when the sun is just above the satellite.
Therefore, the maximum albedo heat input is $\dot{Q}_\mathrm{a} =
2a\,\dot{Q}_\mathrm{s}\,,$ where $a$ denotes the average albedo coefficient.
The planetary IR irradiation heat input is given by
$$\dot{Q}_\mathrm{p} = (A/2)\e\s\,T_\mathrm{p}^4\,,$$ 
where $T_\mathrm{p}$ is the planet's (Earth's) blackbody-equivalent
temperature, the projected area is a half of the real area (like for albedo
absorption), and the satellite's surface IR absorptivity is taken equal to its
IR emissivity $\e$.  The value of $T_\mathrm{p}$ derives from the planet's
heat balance equation \cite{Kreith} 
$$
4\s T_\mathrm{p}^4 = (1-a)\,G_\mathrm{s}\,.
$$
The standard value of the average albedo coefficient of the Earth is $a=0.3$,
which gives $\s\,T_\mathrm{p}^4 = 239.7\;\mathrm{W}\mathrm{m}^{-2}$ (using the
value of $G_\mathrm{s}$ in Table~\ref{tab1}).%
\footnote{The corresponding value of the Earth blackbody-equivalent
temperature is $T_\mathrm{p}=255$ K.} 
In conclusion, the IR irradiation heat input can be computed with the formula: 
\begin{equation}
\dot{Q}_\mathrm{p} =  \e A \, (119.9\;\mathrm{W}\mathrm{m}^{-2}).
\label{Qp}
\end{equation}

P\'erez-Grande et al's two-node thermal model is applicable to
micro-satellites. Specifically, we have in mind a micro-satellite with the
shape of a cube of $0.5$~m side and with mass of about 50~kg, mostly covered
by solar cells.  The relevant values of the satellite and orbital parameters
for two model examples are collected in Table~\ref{tab1} and employed in
Sects.~\ref{num} and \ref{num-sol}.

We can write equations (\ref{dimODE_1}) and (\ref{dimODE_2}) in a
non-dimensional form by defining 
{\setlength\arraycolsep{2pt}
\begin{eqnarray*}
b &=& A\e\s/(C_\mathrm{e}\nu), \\
q_\mathrm{s} &=& b^{1/3}\,\dot{Q}_\mathrm{s}/(C_\mathrm{e}\nu), \quad
q_\mathrm{a} = b^{1/3}\,\dot{Q}_\mathrm{a}/(C_\mathrm{e}\nu), \\
q_\mathrm{p} &=& b^{1/3}\,\dot{Q}_\mathrm{p}/(C_\mathrm{e}\nu), \quad
q_\mathrm{i} = b^{1/3}\,\dot{Q}_\mathrm{i}/(C_\mathrm{e}\nu), \\
k &=& K_\mathrm{ie}/(C_\mathrm{e}\nu), \quad
r = R_\mathrm{ie}/(C_\mathrm{e}\nu\,b), \\
c &=& C_\mathrm{i}/C_\mathrm{e}\,,
\end{eqnarray*}
}%
and defining non-dimensional temperature variables
$$\T_\mathrm{e}=b^{1/3}\,T_\mathrm{e}\,,\;
\T_\mathrm{i}=b^{1/3}\,T_\mathrm{i}\,,$$ 
and a time variable $\nu t$ (which is still denoted by $t$ for
notational simplicity).%
\footnote{Notice that the non-dimensionalization procedure is similar to 
the one used in Ref.~\cite{NoDy} but the notation is somewhat different: 
$k$ refers now to the thermal conductance while the non-dimensional heat
inputs are denoted by $q$ with the corresponding subscript; and $b$
denotes the constant that before was named $a$ (since $a$ is now reserved for 
the albedo coefficient).}  
Thus, we obtain the non-dimensional ODE's
{\setlength\arraycolsep{2pt}
\begin{eqnarray}
\dot{\T}_\mathrm{e} &=& q_\mathrm{p} + q_\mathrm{s}\, f_s(t) +
q_\mathrm{a}\,f_a(t) 
+ k (\T_\mathrm{i} - \T_\mathrm{e}) + r (\T_\mathrm{i}^4 - \T_\mathrm{e}^4)  
- \T_\mathrm{e}^4,
\label{ODE_1}\\
c\,\dot{\T}_\mathrm{i} &=& q_\mathrm{i}
+ k (\T_\mathrm{e} - \T_\mathrm{i}) + r (\T_\mathrm{e}^4 - \T_\mathrm{i}^4).
\label{ODE_2}
\end{eqnarray}
}%
This system of two non-autonomous nonlinear ODE's is difficult to solve.
As a first step, we remove the oscillating terms by averaging
$f_\mathrm{s}$ and $f_\mathrm{a}$, like in Ref.~\cite{NoDy}.

\subsection{Steady state temperatures}
\label{steady}

If $q_\mathrm{e}$ denotes the time average of the external heat input
$q_\mathrm{p} + q_\mathrm{s}\, f_\mathrm{s}(t) + q_\mathrm{a}\,
f_\mathrm{a}(t)$, then the averaging of the dynamical equations (\ref{ODE_1})
and (\ref{ODE_2}) results in the autonomous system:
{\setlength\arraycolsep{2pt}
\begin{eqnarray}
\dot{\T}_\mathrm{e} &=& q_\mathrm{e} 
+ k (\T_\mathrm{i} - \T_\mathrm{e}) + r (\T_\mathrm{i}^4 - \T_\mathrm{e}^4)  
- \T_\mathrm{e}^4,
\label{avODE_1}\\
c\,\dot{\T}_\mathrm{i} &=& q_\mathrm{i}
+ k (\T_\mathrm{e} - \T_\mathrm{i}) + r (\T_\mathrm{e}^4 - \T_\mathrm{i}^4).
\label{avODE_2}
\end{eqnarray}
}%
Instead of the one-node model averaged equation \cite{NoDy}, which is
straightforward to solve, we have now a system of {\em two} autonomous
nonlinear ODE's. The standard way of analyzing these systems begins with the
localization of their fixed points \cite{Andro}. 

The fixed points of Eqs.~(\ref{avODE_1}) and (\ref{avODE_2}) are the solutions
of two fourth degree algebraic equations.  These two equations can be solved
by first eliminating one unknown, namely, $\T_\mathrm{i}$, as is easily done
by adding both equations, which yields $\T_\mathrm{e}^4 = q_\mathrm{e} +
q_\mathrm{i}$.  Thus, the outer node temperature is given by $\T_\mathrm{e} =
(q_\mathrm{e} + q_\mathrm{i})^{1/4}$.  It only depends on the total heat
input, like the steady state temperature in the one-node model \cite{NoDy}.
Once $\T_\mathrm{e}$ is known, we can substitute for it in one of the
algebraic equations, say the second one, to obtain a fourth degree algebraic
equation for $\T_\mathrm{i}\,,$ namely, 
\begin{equation}
\label{eq4qi}
q_\mathrm{i} + k \T_\mathrm{e} + r \T_\mathrm{e}^4 - k \T_\mathrm{i} -
r\T_\mathrm{i}^4 = 0.
\end{equation}
According to the fundamental theorem of algebra \cite{Tignol}, this equation
has four complex roots (solutions), of which an even number (0, 2 or 4) are
real.  We are interested in just the real positive roots.  To this end, we
apply Descartes's rule of signs, which says that no equation can have more
positive roots than it has changes of signs in the coefficients
\cite{Hymers}. Clearly, there is only one change of sign, so there is one
positive root at the most. If we replace $\T_\mathrm{i} \ra -\T_\mathrm{i}$ in
the equation, we also deduce that there is one negative root at the
most. Therefore, the equation can have a positive and a negative root or no
real roots at all. Given that the corresponding polynomial is positive for
$\T_\mathrm{i}=0$ and becomes negative as $\left|\T_\mathrm{i}\right| \ra
\infty$, there is at the least one real root. Thus, the equation must have a
positive and a negative root, in addition to a pair of complex conjugate
roots.

The equation $\T_\mathrm{e}^4 = q_\mathrm{e} + q_\mathrm{i}$ also has a
positive root, a negative root, and a pair of complex conjugate roots, but
these are just the four complex fourth roots of a positive real number.  When
we consider the two fourth-degree equations together, the total number of
roots is $4\times 4 = 16$, but only one has positive $\T_\mathrm{e}$ and
$\T_\mathrm{i}$ .  The positive root of Eq.~(\ref{eq4qi}) has a complicated
expression in terms of radicals of the coefficients of the equation
\cite{Tignol,Hymers}. Assuming that the coefficients have numerical values,
it is much more convenient to find the root by numerical methods.

The existence of one and only one couple of positive steady-state temperatures
is, of course, in accord with our physical intuition. In fact, one may wonder
why there are other real solutions with negative absolute temperatures and why
the flow given by Eqs.~(\ref{avODE_1}) and (\ref{avODE_2}) crosses the axes
$\T_\mathrm{e}=0$ or $\T_\mathrm{i}=0$.  In this regard, let us notice that
those equations are not valid when $\T_\mathrm{e}$ or $\T_\mathrm{i}$ vanish,
because the thermal capacities $C_\mathrm{e}$ and $C_\mathrm{i}$ can only be
taken constant in an interval of temperatures, which is usually long but
cannot be extended to zero absolute temperature. In fact, the third law of
thermodynamics implies that any thermal capacity vanishes as the absolute
temperature approaches zero \cite{Callen}.

\subsection{Stability of the steady state}
\label{stab}

To determine the stability of the steady state we have to calculate the
Jacobian matrix of the vector field defined by the ODE's
\cite{Drazin,Andro,Hi-Sm}, namely, of the vector field
\begin{equation}
\label{vector}
\left\{q_\mathrm{e} + k (\T_\mathrm{i} - \T_\mathrm{e}) + r (\T_\mathrm{i}^4 -
\T_\mathrm{e}^4) - \T_\mathrm{e}^4\,,\;
c^{-1} \left[q_\mathrm{i} + k (\T_\mathrm{e} - \T_\mathrm{i}) + r
(\T_\mathrm{e}^4 - \T_\mathrm{i}^4)\right]\right\}. 
\end{equation}
If the Jacobian matrix is nonsingular, the fixed point is simple and the signs
of the eigenvalues of the Jacobian matrix determine the nature and stability
of the fixed point. In particular, if both eigenvalues are negative or, more
generally, have negative real parts, the fixed point is an asymptotically
stable sink, at least, locally.

The Jacobian matrix 
\begin{equation}
\label{Jacob}
J = \left(
\begin{array}{cc}
- k - 4 (r+1)  \T_\mathrm{e}^3  &  k + 4 r  \T_\mathrm{i}^3  \\
\left(k + 4 r \T_\mathrm{e}^3\right)/c  &  
\left(- k - 4 r  \T_\mathrm{i}^3\right)/c  
\end{array}  \right).
\end{equation}
is indeed nonsingular, for
$$\det J = J_{11}J_{22} - J_{12}J_{21} = \frac{4 \,\T_\mathrm{e}^3\left(k + 4
r \T_\mathrm{i}^3\right)}{c} > 0.$$
The eigenvalue equation 
$$
(J_{11}-\l)(J_{22}-\l) - J_{12}J_{21} = 
J_{11}J_{22} - J_{12}J_{21} - (J_{11}+J_{22}) \l  + \l^2 = 0
$$
has discriminant 
\begin{equation}
\label{discrim}
\Delta = (J_{11}+J_{22})^2 - 4 (J_{11}J_{22} - J_{12}J_{21}) = 
(J_{11}-J_{22})^2 + 4 J_{12}J_{21}\,. 
\end{equation}
It is positive, given that $J_{12}J_{21} >0$.  Therefore, both eigenvalues are
real and different. This implies that the matrix $J$ is diagonalizable. The
larger eigenvalue 
$$
\l = \frac{1}{2}\left(J_{11}+J_{22} + \sqrt{\Delta} \right)
$$
is negative if $\sqrt{\Delta} < -(J_{11}+J_{22})$, that is to say, if
$\Delta < (J_{11}+J_{22})^2$, equivalent to $J_{11}J_{22} - J_{12}J_{21} > 0$.
In consequence, both eigenvalues are negative, so the fixed point is a stable
sink and, to be specific, it is a node.

The local asymptotic stability of the fixed point is, again, in accord with
our physical intuition. Furthermore, we expect that the fixed point be a {\em
global} sink in the positive temperature quadrant.  This can be proved with
the help of Bendixson's criterion \cite{Andro}:
if on a simple connected region the divergence of the ODE's vector field does
not change sign, then the ODE's has no closed trajectories lying entirely in
that region.  This criterion is applicable, for the divergence equals
$J_{11}+J_{22}$, which is always negative in the positive temperature
quadrant.  The absence of closed trajectories must be combined with the
Poincar\'e-Bendixson theorem \cite{Andro,Hi-Sm,Drazin}, which restricts the
generic behavior of an autonomous system of two first-order ODE's to having
``simple'' attracting sets, namely, fixed points or limit cycles.  On account
of the trivial fact that the flow points towards the interior of the positive
temperature quadrant, we conclude that the flow must end at its unique sink in
the quadrant, which is therefore globally stable in it.

\subsection{Numerical examples}
\label{num}

After establishing that the flow has a globally stable sink, it is useful to
see the aspect of the flow for sensible values of the parameters by plotting
the ODE's vector field (\ref{vector}).  
In addition, the computation of 
the eigenvalues and eigenvectors of the respective Jacobian matrices provides
a very precise local picture of the flow around the fixed points.
that confirms and completes the global picture provided by the vector
fields.
We examine two examples: one in which the
two nodes are strongly coupled (large values of $K_\mathrm{ie}$ and
$R_\mathrm{ie}$) and another in which the coupling is weak (small
$K_\mathrm{ie}$ and $R_\mathrm{ie}$). The corresponding values of all the
parameters are given in Table~\ref{tab1}.%
\footnote{The numerical values are assumed to have four digit precision, 
even when that is not the number of digits explicitly shown: 
then, the remaining digits are zeros.}

\begin{table}
\begin{center}
\begin{tabular}{p{8cm}ll}
 &  Example 1 & Example 2 \\
\hline\\[-3mm]
Outer node area $A$ (m$^2$) &  $1.5$ & idem \\
Outer node heat capacity $C_\mathrm{e}$ (J\hspace{2pt}K$^{-1}$)  & 
18\hspace{1pt}000 & idem  \\
Inner node heat capacity $C_\mathrm{i}$ (J\hspace{2pt}K$^{-1}$)  & 
13\hspace{1pt}500 & 24\hspace{1pt}000  \\
Conductance $K_\mathrm{ie}$ (W\hspace{2pt}K$^{-1}$)     &  10  & 1 \\
Radiative coupling constant $R_\mathrm{ie}$ (W\hspace{2pt}K$^{-4}$) &
$3.6\cdot 10^{-8}$  & $9\cdot 10^{-9}$ \\
Solar absorptivity $\a_\mathrm{s}$  &  0.8 & idem \\
Emissivity $\e$  & 0.7 & idem \\
Internal heat dissipation (W)  &  40 & 80 \\
\hline\\[-3mm]
Solar constant $G_\mathrm{s}$ (W\hspace{2pt}m$^{-2}$) &  1\hspace{1pt}370 & 
idem \\
Earth's albedo coefficient $a$  &  0.3 & idem \\
Orbital frequency $\nu$ &  (1.5~h)$^{-1}$ & idem \\
Solar time fraction  $2 x_1$ &  0.6 & idem \\
Albedo time fraction  $2 x_2$ & 0.5 & idem \\
\hline
\end{tabular}
\end{center}
\caption{Satellite and orbital parameters for two numerical examples.}
\label{tab1}
\end{table}

\subsubsection{Example 1: strong coupling}
\label{Example1}

From the parameters in Table~\ref{tab1}, we can obtain the non-dimensional
parameters as follows. 
We first calculate the denominator
$C_\mathrm{e}\,\nu = 18\hspace{1pt}000\,\mathrm{JK}^{-1}\, (1.5
\cdot 3600 \,\mathrm{s})^{-1} = 3.333$ W\hspace{2pt}K$^{-1}$, and then
calculate:
{\setlength\arraycolsep{2pt}
\begin{eqnarray*}
b &=& \frac{1.5 \cdot 0.7 \cdot 5.67 \cdot 10^{-8}}{3.333}
\,\mathrm{K}^{-3} = 1.786\cdot 10^{-8} \,\mathrm{K}^{-3}, \\
k &=& \frac{10}{3.333} = 3, \quad r = \frac{3.6\cdot
10^{-8}\,\mathrm{K}^{-3}}{3.333\,b} = 0.6047,\\ q_\mathrm{s} &=&
b^{1/3}\frac{1370\cdot 0.8\cdot 1.5/4}{3.333 \,\mathrm{K}^{-1}} = 0.3223,
\quad q_\mathrm{i} = \frac{40\,b^{1/3}}{3.333 \,\mathrm{K}^{-1}} = 0.03137,
\quad
c = \frac{13\hspace{1pt}500}{18\hspace{1pt}000} = 0.75.
\end{eqnarray*}
}%
To calculate $q_\mathrm{e}$ we need the average of the periodic functions
$f_\mathrm{s}$ and $f_\mathrm{a}$, which yields $q_\mathrm{e} =
q_\mathrm{p} + 2x_1 q_\mathrm{s} + q_\mathrm{a}/\pi$
\cite{NoDy}.
We have that $q_\mathrm{a} = 2 a q_\mathrm{s} = 0.6\cdot 0.3223 = 0.1934$,
and, on account of Eq.~(\ref{Qp}),
$$q_\mathrm{p} = \frac{b^{1/3}(0.7\cdot 1.5\cdot
119.9)}{3.333\,\mathrm{K}^{-1}} =  0.09872,$$
so that
$$q_\mathrm{e} = 0.09872 + 0.6\cdot 0.3223 + 0.1934/\pi = 0.3537.$$

\begin{figure}
\centering{\includegraphics[width=8cm]{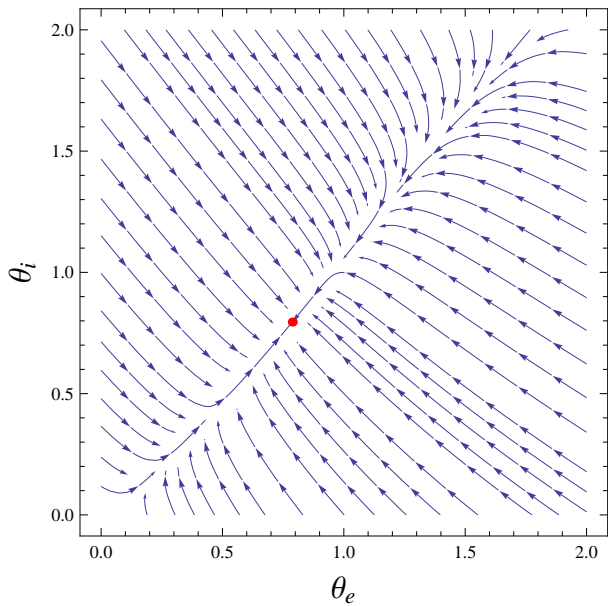}}
\caption{
Temperature flow of a model with strong coupling. The steady
state is marked by a dot.  Multiplying $\T_\mathrm{e}$ and $\T_\mathrm{i}$ by
$b^{-1/3} = 382.6$~K one obtains the absolute temperatures $T_\mathrm{e}$ and
$T_\mathrm{i}$.}
\label{flow1}
\end{figure}

With the given values of the non-dimensional parameters, we can compute the
fixed point by using, for example, Newton's method, with initial values
$\T_\mathrm{e} = \T_\mathrm{i} = 1$. The solution is $\T_\mathrm{e}^* =
0.7877, \;\T_\mathrm{i}^* = 0.7952$ (to which correspond $T_\mathrm{e}^* =
301.4~\mathrm{K}, \;T_\mathrm{i}^* = 304.2~\mathrm{K}$).  We can also compute
the eigenvalues of the Jacobian matrix at
$\left(\T_\mathrm{e}^*,\T_\mathrm{i}^*\right)$, which must be negative,
according to Sect.~\ref{stab}.  The computation of the eigenvalues yields
$\{-10.74, -1.024\}$, which are indeed negative; besides, they are very
different in absolute value, due to the magnitude of the discriminant $\Delta$
(\ref{discrim}). Therefore, the flow converges to the node much faster in the
direction of the eigenvector corresponding to the eigenvalue $-10.74$ than in
the direction of the eigenvector corresponding to $-1.024$.  The former
eigenvector is $(-0.6759, 0.7370)$ (or a multiple thereof), whereas the latter
is $(0.6362, 0.7716)$.  The consequent local flow pattern is borne out by the
plot of the vector field flow in the square $[0,2] \times [0,2]$ that is shown
in Fig.~\ref{flow1}.

The qualitative features of the flow can be deduced from the results in
Sects.~\ref{steady} and \ref{stab}. As the coupling parameters $k$ and $r$ are
large in comparison with the heat input $q_\mathrm{i}$, we can make
$q_\mathrm{i}=0$ in Eq.~(\ref{eq4qi}) for $\T_\mathrm{i}$, which implies that
$\T_\mathrm{i} = \T_\mathrm{e}$.  Notice that any $q_\mathrm{i} > 0$ gives
rise to $\T_\mathrm{i} > \T_\mathrm{e}$, as is natural on physical grounds.
Thus, we put $\T_\mathrm{i} = \T_\mathrm{e} + \d$ in Eq.~(\ref{eq4qi}), expand
it to first order in $\d$, and solve the resulting linear equation for $\d$ to
obtain
$$
\d = \frac{q_\mathrm{i}}{k + 4r\T_\mathrm{e}^3} \,.
$$
The resulting numerical value $\d \simeq 0.0075$ coincides with the
difference $\T_\mathrm{i}^* - \T_\mathrm{e}^*$ found in the numerical
computation.
As regards the Jacobian matrix $J$ and the discriminant $\Delta$, given by
Eqs.~(\ref{Jacob}) and (\ref{discrim}), respectively, we can make in these
equations $\T_\mathrm{i} = \T_\mathrm{e} = 0.7877$. 
Thus, the discriminant
$$
\Delta = (J_{11}-J_{22})^2 + 4 J_{12}J_{21} 
\simeq 93.6. 
$$
This yields a difference between eigenvalues $\sqrt{\Delta} \simeq 9.7$, which
is very close to the already found value. Of course, the large value of
$\sqrt{\Delta}$ is the cause of the appearance of a ``fast variable'' and a
``slow variable'', such that the flow is first attracted to the almost
diagonal curve in Fig.~\ref{flow1}, along which it flows towards the fixed
point. Therefore, the temperatures of the two nodes quickly become
approximately equal and then the common temperature evolves towards its steady
value. This type of evolution justifies the isothermal model studied in
Ref.~\cite{NoDy}.

\subsubsection{Example 2: weak coupling}
\label{Example2}

As a second example, we consider smaller values of the thermal couplings
between nodes.  In practice, the conductance $K_\mathrm{ie}$ can be
substantially reduced through a reduction of the joints between the
satellite's core and its outer shell, in addition to the use of thermally
insulating material. The radiative coupling $R_\mathrm{ie}$ can also be
reduced by using materials with low absorptivities and emissivities for the
relevant surfaces.  The new values of $K_\mathrm{ie}$ and $R_\mathrm{ie}$ are
displayed in the last column of Table~\ref{tab1}.  On the other hand, as seen
in the table, we assume larger internal heat capacity and dissipation, to
enhance the differences with the preceding example.

Among the five non-dimensional parameters entering in Eqs.~(\ref{avODE_1}) and
(\ref{avODE_2}), the non-dimensional external heat input keeps its former
value, 
namely, $q_\mathrm{e}=0.3537$, but $k,\, r,
\,q_\mathrm{i}$ and $c$ adopt new values, namely,
\begin{eqnarray*}
k = 0.3, \quad r = 0.1512, \quad q_\mathrm{i} = 0.06274, \quad c = 1.333.
\end{eqnarray*} 
Using these values, we compute the fixed point (with Newton's method and
starting with the same initial values $\T_\mathrm{e} = \T_\mathrm{i} = 1$); we
obtain $\T_\mathrm{e}^* = 0.8033, \;\T_\mathrm{i}^* = 0.8966$ (to which
correspond $T_\mathrm{e}^* = 307.3~\mathrm{K}, \;T_\mathrm{i}^* =
343.0~\mathrm{K}$).  We obtain the eigenvalues $\{-2.835, -0.4036\}$ and the
respective eigenvectors $(-0.9803, 0.1975)$ and $(0.3067, 0.9518)$. The
eigenvalues are not as different in absolute value as in the preceding example
(now $\sqrt{\Delta} = 2.432$). Nevertheless, the convergence is still
considerably slower in the $(0.3067, 0.9518)$ direction.

\begin{figure}
\centering{\includegraphics[width=8cm]{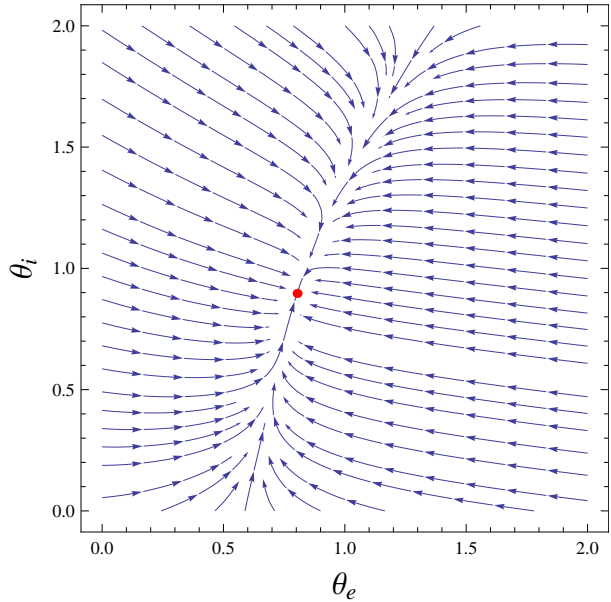}}
\caption{Temperature flow of a model with weak coupling. 
Multiplying $\T_\mathrm{e}$ and $\T_\mathrm{i}$ by $382.6$~K
one obtains $T_\mathrm{e}$ and $T_\mathrm{i}$.}
\label{flow2}
\end{figure}

The flow in the square $[0,2] \times [0,2]$ is shown in Fig.~\ref{flow2}.
Since there is a relatively fast variable, the flow is again initially
attracted to a curve; but now this curve, which approximately goes along the
eigenvector $(0.3067, 0.9518)$, is not close to the diagonal. In fact, for
some initial values of the temperatures, the difference between the two
temperatures oscillates, taking both signs (the trajectory crosses the
diagonal).  Notice that the reduction to an isothermal model is not
appropriate in the weakly coupled case.

We remark that it is possible to reduce further the value of the discriminant
$\Delta$ while keeping $q_\mathrm{e} > 0,\, q_\mathrm{i} > 0,\, k> 0,\, r> 0$
and $c>0.$ In fact, keeping $q_\mathrm{e} = 0.3537$, one can get $\Delta
\simeq 0.02$ for small values of the other parameters (especially, $k$ and $r$
but also $c$ and $q_\mathrm{i}$).  Thus, there are no fast and slow
variables. However, those small values of $k,\, r,\, c$ and $q_\mathrm{i}$
correspond to hardly realizable values of the physical parameters.

\subsection{Driven two-node model}
\label{forced}

In this section, we study the original Eqs.~(\ref{ODE_1}) and (\ref{ODE_2}),
without averaging the periodic driving term
$q_\mathrm{p} + q_\mathrm{s}\, f_\mathrm{s}(t) +
q_\mathrm{a}\,f_\mathrm{a}(t).$ 
As in Ref.~\cite{NoDy}, it is convenient to redefine this term as
\begin{equation}
\label{f}
f(t) = q_\mathrm{p} + q_\mathrm{s}\, f_\mathrm{s}(t) +
q_\mathrm{a}\,f_\mathrm{a}(t) - q_\mathrm{e}\,,
\end{equation}
such that it has vanishing average over a period and represents the deviations
about the mean.  
The addition of $f(t)$ converts the autonomous system of
the two ODE's (\ref{avODE_1}) and (\ref{avODE_2}) in an autonomous system of
{\em three} ODE's, the third equation being $\dot{t} = 1$ (the new system is
called the {\em suspended} system).
Three-dimensional autonomous systems can have very complex flows
as is well known; in particular, they 
and can have {\em chaotic attractors} \cite{Gu-Ho,Drazin}.

Complex three-dimensional flows, in particular, chaotic flows, can arise from
simpler flows through instabilities and bifurcations.  
Drazin \cite{Drazin}
distinguishes four routes to chaos, three of which can operate in our case.
(i) subcritical instability, (ii) a sequence of bifurcations, (iii) period
doubling, and (iv) intermittent transition. They all have some common
elements, but the second route is not relevant to our problem, for it
requires, according to Drazin's description, a phase space of high dimension
(it is the route that is believed to lead to fluid turbulence).  We cannot
rule out the other three routes.
All the transitions to chaos take place as a parameter of a nonlinear system
is increased and, hence, a simple attractor gives rise to a chaotic
attractor. The transition can occur through a succession of instabilities or
at once, as in the case of a subcritical instability.
The parameter for instability and chaos in our system is the magnitude of 
the driving heat oscillation $f$.
However, if the magnitude of $f$ is not large, 
we can prove the existence of one and only
one attracting limit cycle. 
But let us specify before what kind of limit cycle appears in our system.

Notice that the flow of the three Eqs.~(\ref{avODE_1}), (\ref{avODE_2}) and
$\dot{t} = 1$ (the suspended averaged system) is attracted to the line
described by the fixed point of just Eqs.~(\ref{avODE_1}) and (\ref{avODE_2})
as $t$ goes from $-\infty$ to $\infty$. This line is turned into a cycle if,
taking advantage of the periodicity in $t$, we restrict the flow to $t \in
[0,1)$ and identify the two-dimensional temperature plane at $t=1$ with the
one at $t=0$. The analogous lower dimensional operation is used in
Ref.~\cite{NoDy}. Geometrically, the operation described in Ref.~\cite{NoDy}
amounts to rolling the temperature-time plane $\mathbb{R}^2$ along time into
the cylinder $\mathbb{R} \times S^1$, where the circular time component $S^1$
reflects the periodicity.  The cylinder is best represented by using polar
coordinates in a plane, the angular coordinate being time.  With one more
temperature dimension, the three-dimensional Euclidean space $\mathbb{R}^3$ is
turned into a generalized ``cylinder'' $\mathbb{R}^2 \times S^1$. We can
actually restrict the temperature plane $\mathbb{R}^2$ to the positive
quadrant and represent its product with $S^1$ in three-dimensional space by
using a set of cylindrical coordinates in which time is the angular
coordinate.  In the representations in which time is an angular coordinate,
the limit cycle of the undriven model is just a circle (around which winds the
line described by the fixed point as $t$ goes from $-\infty$ to $\infty$).

In the two-dimensional case of Ref.~\cite{NoDy}, the limit cycle for $f=0$ is
deformed by the driving but still remains an attracting limit cycle.  This
constitutes an example of {\em structural stability} and is proved with
qualitative methods (based on the Poincar\'e-Bendixson-Dulac theory) and also
with a perturbation method.  In three (or more) dimensions, we can only employ
perturbation theory. In fact, the existence and uniqueness of the limit cycle
in a range of the perturbation parameter is a consequence of the averaging
theorem stated by Guckenheimer \& Holmes \cite{Gu-Ho}.  In the next section,
we generalize the perturbation method of Ref.~\cite{NoDy}, which allows us to
compute the limit cycle and, thus, constitutes a constructive proof of its
existence and uniqueness.  The conclusion is that the two-node model behaves
as a sort of driven nonlinear oscillator and can be related to the typical
cases studied in classic textbooks \cite{Nay1,Andro}. This conclusion is valid
in a range of magnitudes of $f$ that is sufficient for realistic applications
(as remarked at the end of Sect.~\ref{num-sol}).

\subsection{Perturbation method}
\label{pertur}

We introduce a formal perturbation parameter $\epsilon$ and 
write Eqs.~(\ref{ODE_1}) and (\ref{ODE_2}) as
{\setlength\arraycolsep{2pt}
\begin{eqnarray}
\dot{\T}_\mathrm{e} &=& q_\mathrm{e} + \epsilon \, f(t)
+ k (\T_\mathrm{i} - \T_\mathrm{e}) + r (\T_\mathrm{i}^4 - \T_\mathrm{e}^4)  
- {\T_\mathrm{e}}^4,
\label{ODE_1-pert}\\
c\,\dot{\T}_\mathrm{i} &=& q_\mathrm{i} + k (\T_\mathrm{e} - \T_\mathrm{i}) +
r (\T_\mathrm{e}^4 - \T_\mathrm{i}^4).
\label{ODE_2-pert}
\end{eqnarray}
}%
Then, we define the vector $\left(\T_\mathrm{e}, \T_\mathrm{i}\right)$ and
assume an expansion of the form
$$
\T_j(t) = \sum_{n=0}^{\infty} \epsilon^n \T_{(n)j}(t)\,,
$$
where $j=\mathrm{e}$ or $\mathrm{i}$.  When we substitute this expansion into
Eqs.~(\ref{ODE_1-pert}) and (\ref{ODE_2-pert}), we obtain at the first order
in $\epsilon$ a couple of linear equations that we can write as
\begin{equation}
\label{linODE-pert}
\dot{\T}_{(1)j} =  \sum_{i} J_{ji}(t) \,{\T}_{(1)i} + F_j(t)\,,
\end{equation}
where the vector $F=(f,0)$ and $J_{ji}(t)$ is the Jacobian matrix
(\ref{Jacob}) calculated at the point $\T_{(0)j}(t)$ that solves the zeroth
order equation (the unperturbed equation).  Eq.~(\ref{linODE-pert}) is to be
solved with the initial condition $\T_{(1)j}(0)=0$.

Since the unperturbed equation is an averaged equation, our perturbation
method can be understood as a method of averaging. Indeed, the natural
solution of an {\em nonhomogeneous} linear equation like
Eq.~(\ref{linODE-pert}) is obtained by variation of parameters \cite{Hur},
which is a simple method of averaging \cite{Nay2}.  Thus, the first step to
solve Eq.~(\ref{linODE-pert}) consists in solving the corresponding
homogeneous equation.  Given an initial condition $\T_{(1)j}(t_0)$, the formal
solution of the homogeneous equation can be expressed in vector form as
\begin{equation}
\T_{(1)}(t) = U(t,t_0) \cdot \T_{(1)}(t_0),
\label{formal_sol}
\end{equation}
where $U(t,t_0)$ is the matrix solution of the homogeneous equation, namely,
$$
\frac{dU}{dt} = J \cdot U,
$$
and $U(t_0,t_0)$ is the identity.  Note that the columns of $U$ are linearly
independent solutions of the homogeneous equation, so its general solution is
a combination of them with arbitrary coefficients. In particular, one can
reinterpret $\T_{(1)}(t_0)$ in Eq.~(\ref{formal_sol}) as a couple of arbitrary
coefficients and $\T_{(1)}(t)$ as the general solution.  One can then find a
solution of the nonhomogeneous equation by assuming that the coefficients are
functions of $t$:
$$
\T_{(1)}(t) = U(t,t_0) \cdot A(t).
$$
Taking the derivative of this equation with respect to $t$, 
substituting in it the derivatives of $\T_{(1)}$ and $U$, and simplifying, 
one obtains an equation for $A(t)$, namely, 
$$
U \cdot \dot{A} = F.
$$
Solving it, the solution of the nonhomogeneous equation with the initial
condition $\T_{(1)j}(0)=0$ is found to be
\begin{equation}
\T_{(1)}(t) = U(t) \cdot \left(\int_0^t U(\t)^{-1}\cdot F(\t) \,d\t\right),
\label{T_1}
\end{equation}
where $U(t) = U(t,0)$. 

This solution can be compared with the solution of the one-dimensional
equation in Ref.~\cite{NoDy}: if we denote by $I(t)$ the corresponding
one-dimensional matrix $U(t)^{-1}$, both expressions coincide. Moreover, the
formula for $I(t)$ given there has a higher dimensional analog:
\begin{equation}
\label{U(J)}
U(t) = \exp\left[\int_0^t J(\t)\, d\t \right]. 
\end{equation}
However, for this formula to be valid, $J$ should commute with its integral,
namely, with the integral in the exponential in that formula \cite{Hur}.
Actually, one must solve the homogeneous equation to find $U$, 
and the solution cannot be reduced to quadratures.%
\footnote{Indeed, the solution of two-dimensional homogeneous equations,
in particular, of second order ODE's with variable coefficients, even simple
ones, gives rise to new functions (Bessel, Hermite, hypergeometric and other
functions), which are only known in terms of their power series. Nevertheless,
many properties of those functions can be deduced from the generating ODE's.}

Let us compare the one-node case \cite{NoDy} with the present two-node case:
in the former case, $\T_{(1)}(t)$ can be expressed in terms of quadratures of
known (but complicated) functions, namely, of $f(t)$ and $\T_{(0)}(t)$, but,
in the latter case, there is no expression in terms of quadratures and,
moreover, the explicit expression of $\T_{(0)}(t)$ is not available.
Nevertheless, following the procedure in Ref.~\cite{NoDy}, we look for
asymptotic expressions valid for long times. Then, $\T_{(0)}(t)$
approaches its fixed point $\T^*$ and the Jacobian matrix tends
to the corresponding limit. Therefore, formula (\ref{U(J)}) is applicable and
Eq.~(\ref{T_1}) becomes
\begin{equation}
{\T}_{(1)}(t) = \int_0^t \exp\left[(t-\t) J \right] \cdot F(\t) \,d\t =
\int_0^t  \exp\left[\t J \right] \cdot F(t-\t) \,d\t.
\label{T_1_Jcst}
\end{equation}
Naturally, this is the solution of Eq.~(\ref{linODE-pert}) with constant $J$,
which is a linear ODE system with constant coefficients.  A linear ODE system
with constant coefficients is solved by finding the eigenvalues and
eigenvectors of the coefficient matrix \cite{Hur}.  Given that the eigenvalues
of $J$ are two different real numbers, $J$ is diagonalizable.  Furthermore,
the eigenvalues are negative, so that the ODE system is equivalent to the
equation of a driven overdamped linear oscillator.  The evolution of the
temperatures consists of a transient part, which depends on the initial
conditions but decays exponentially, and a periodic part, which is independent
of the initial conditions and represents the limit cycle (at the first
perturbative order).  The periodic part can be obtained by extending the upper
integration limit of the last integral in Eq.~(\ref{T_1_Jcst}) from $t$ to
$\infty$: 
\begin{equation}
{\T}_{(1)}^\infty(t) = \int_0^\infty \exp\left[\t J \right] \cdot F(t-\t)
\,d\t.
\label{T_1_lim}
\end{equation}
It is convenient to express this formula in the basis of the eigenvectors of
$J$, but only for numerical calculations, because the analytical expressions
of the eigenvalues and eigenvectors of $J$ in terms of the parameters
($q_\mathrm{e},\,q_\mathrm{i},\, k,\, r$ and $c$) are cumbersome. 

Given that ${\T}_{(1)}^\infty(t)$ is a periodic function, it can be expanded
in a Fourier series.
This is done by inserting the Fourier series of $F(t)$ in the integral of
Eq.~(\ref{T_1_lim}) and integrating term by term.  Alternately, we can solve
Eq.~(\ref{linODE-pert}) by substituting into it the Fourier series of both
${\T}_{(1)}^\infty(t)$ and $F(t)$ and then solving for the Fourier
coefficients of ${\T}_{(1)}^\infty(t)$. The result is
\begin{equation}
{\T}_{(1)}^\infty(t) = \sum_{m = -\infty}^{\infty} 
e^{2\pi i m t} \left(2\pi i m I - J\right)^{-1} \cdot F(m)\,, 
\label{Fourier-sol}
\end{equation}
where $I$ is the $2 \times 2$ identity matrix and $F(m)$ are the Fourier
coefficients of $F(t)$.  For numerical work, this formula can be conveniently
expressed in the basis of the eigenvectors of $J$, like formula
(\ref{T_1_lim}).  For eigenvalues of such a small magnitude that the integrand
of Eq.~(\ref{T_1_lim}) decreases too slowly with $\tau$,
Eq.~(\ref{Fourier-sol}) is preferable. But Eq.~(\ref{T_1_lim}) is generally
suitable for both numerical and analytical work.  In particular, it is
suitable for analyzing the convergence of the perturbative method, as we do
next.

\subsubsection{The perturbation method at higher order}
\label{high-order}

The accuracy of a first order calculation in perturbation theory depends on
the convergence properties of the perturbative expansion. The simplest test of
convergence consists in calculating the second order and comparing it with the
first one. 

The calculation of the second order equation yields
\begin{equation}
\label{linODE-pert2}
\dot{\T}_{(2)j} = \sum_{i} J_{ji}(t) \,{\T}_{(2)i} + \frac{1}{2} \sum_{kl}
H_{j,kl}(t) \,{\T}_{(1)k}(t) \,{\T}_{(1)l}(t)\,,
\end{equation}
where $H_{j,kl}(t)$ is the second derivative (Hessian) tensor of the vector
field of Eqs.~(\ref{ODE_1-pert}) and (\ref{ODE_2-pert}) calculated at
the point $\T_{(0)j}(t)$. 
Although the second order equation seems more complex than the first order
one, it is also a nonhomogeneous linear vector ODE. The only difference is
that the driving $F_j$ is replaced with
$$\widetilde{F}_j = \frac{1}{2}\sum_{kl} H_{j,kl} \,{\T}_{(1)k} \,{\T}_{(1)l}\,,$$
which is also a known function, assuming that the first order equation is
already solved.  Furthermore, the initial condition for
Eq.~(\ref{linODE-pert2}) is, likewise, $\T_{(2)j}(0)=0$.  Therefore, the
nonhomogeneous linear ODE solution (\ref{T_1}) and the expression of the limit
cycle (\ref{T_1_lim}) hold after replacing $F$ with $\widetilde{F}$.

Since the nonhomogeneous linear ODE solution given by Eq.~(\ref{T_1}) is
proportional to the driving heat input, to compare the first and second order
terms of the perturbative expansion, we only need to compare the respective
driving terms. This comparison is not easy in general, but we can easily
compare their perturbative contributions to the limit cycle.  To do this, we
need a property of contractive operators: let $A$ be a matrix with eigenvalues
that have negative real parts, in particular, the matrix of a linear ODE
system with a sink; then, there are constants $k > 0,\; b >0$ such that
$$
\left| e^{t A}\cdot x \right| \leq k\, e^{-t b} \left| x \right|
$$
for all $t \geq 0$ and $x$ (theorem 1 of chapter 7 in Ref.~\cite{Hi-Sm}). 
From Eq.~(\ref{T_1_lim}), and using this property,
$$
\left| {\T}^{\infty}_{(1)}(t) \right| \leq 
\int_0^\infty \left|e^{\t J} \cdot F(t-\t)  \right| d\t
\leq \int_0^\infty k_J \,e^{-\t b_J} \left| F(t-\t)  \right| d\t
\leq m_f \, \frac{k_J}{b_J}\,,
$$
where $m_f = \max_{t}\left|F(t)\right| = \max_{t}\left|f(t)\right|$ and $k_J$
and $b_J$ are constants, once the matrix $J$ is given.  Of course, there is a
similar bound for $\left| {\T}^{\infty}_2(t) \right|$ with $\widetilde{F}$
instead of ${F}$ .  Notice that $\widetilde{F}$ is proportional to the square
of ${\T}_{(1)}$, which shows that $\left|{\T}^{\infty}_{(1)} \right|$ 
is required to be small. In turn, $\left| {\T}^{\infty}_{(1)}
\right|$ is small if $m_f$ is.  In conclusion, the crucial condition for
convergence of the perturbation series (up to the second order) is a
sufficiently small amplitude of the driving heat oscillation.

The preceding conclusion can be extended to higher orders of perturbation
theory. Indeed, the $n$-th order perturbative equation is also an
nonhomogeneous linear vector ODE in which the homogeneous part is given by the
Jacobian matrix $J$ and the driving term is a combination of lower order
solutions, namely, ${\T}_{(m)}$ for $1 \leq m \leq n-1$.  One can express all
these solutions ${\T}_{(m)}$ in terms of just ${\T}_{(1)}$ or the driving $F$,
if one so wishes.  However, one must take into account that the number of
terms involved in the $n$-th order driving term, that is, the number of
combinations of ${\T}_{(1)}$ (or $F$) involved, grows rapidly with $n$.  This
growth could hinder the convergence of the perturbative series.  The relevant
combinatorial factors are independent of the dimension and, therefore, the
argument for the convergence of the series in the one-node case \cite{NoDy}
still holds (the argument is based on a graphical analysis in terms of
``rooted trees'').  Although the perturbative series converges for
sufficiently small amplitude of $f$, the effective calculation of the bound to
this amplitude would now be even harder that in the one-node case.

\subsection{Numerical solutions of the equations with driving}
\label{num-sol}

It is useful to compute a few numerical solutions of Eqs.~(\ref{ODE_1}) and
(\ref{ODE_2}) to see how they converge to the limit cycle.  We use a classical
fourth-order Runge-Kutta method with step-size $1/100$ and select the
parameters values of Example 1 (Table~\ref{tab1}), which yield
{\setlength\arraycolsep{2pt}
\begin{eqnarray*}
q_\mathrm{s} &=& 0.3223,
\quad 
q_\mathrm{p} =  0.09872,
\quad 
q_\mathrm{a} = 2 a q_\mathrm{s} = 0.1934,\\
q_\mathrm{i} &=&  0.03137,
\quad
k = 3, \quad r = 0.6047,
\quad 
c = 0.75.
\end{eqnarray*}
}%
We can take advantage of the results in Subsect.~\ref{Example1} for the
corresponding averaged equations.  Their fixed point (see Fig.~\ref{flow1})
becomes a straight line in $\mathbb{R}^3$ when we add the time dimension.  The
driving $f(t)$ deforms this line into a curve (consisting of repetitions of
the limit cycle).  To integrate the equations with driving, it is convenient
to choose initial conditions that are in a neighborhood of that fixed point.

\begin{figure}
\centering{\includegraphics[width=8cm]{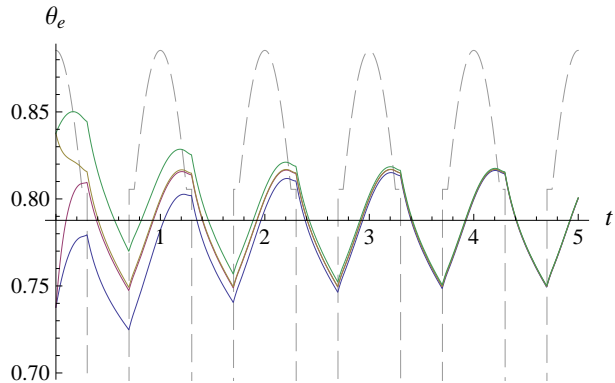}}
\caption{Numerical integration of the driven model
corresponding to Example 1, showing the convergence of $\T_\mathrm{e}(t)$ to
the limit cycle.  The four initial conditions correspond to pairing two
different values of $\T_\mathrm{e}$ with two of $\T_\mathrm{i}$.
The dashed line represents the temperature equivalent driving
$\left[q_\mathrm{e} + f(t)\right]^{1/4}$.}
\label{t-evol_2node}
\end{figure}

We choose as initial conditions nine points $(\T_\mathrm{e},\T_\mathrm{i})$
placed on a $3 \times 3$ grid centered on the fixed point, with spacing of
$0.05$ between points.  To visualize the integral curves, we must find
suitable representations of them.  Unlike in the case of the averaged
equations, the representation in the plane $(\T_\mathrm{e},\T_\mathrm{i})$ is
inadequate, for the curves cross. A two-dimensional representation is possible
by selecting one temperature and plotting its time evolution, like was done
with the only temperature of the one-node model in Ref.~\cite{NoDy}. In fact,
a useful comparison with the one-node model is provided by selecting the outer
node temperature $\T_\mathrm{e}$. The plot of $\T_\mathrm{e}(t)$ for the four
initial conditions defined by the corners of the $3 \times 3$ grid is
displayed in Fig.~\ref{t-evol_2node}, which can be compared with Fig.~1 of
Ref.~\cite{NoDy} (the dashed line in Fig.~\ref{t-evol_2node} also stands for
the temperature equivalent driving $\left[q_\mathrm{e} + f(t)\right]^{1/4}$).

\begin{figure}
\centering{\includegraphics[width=7cm]{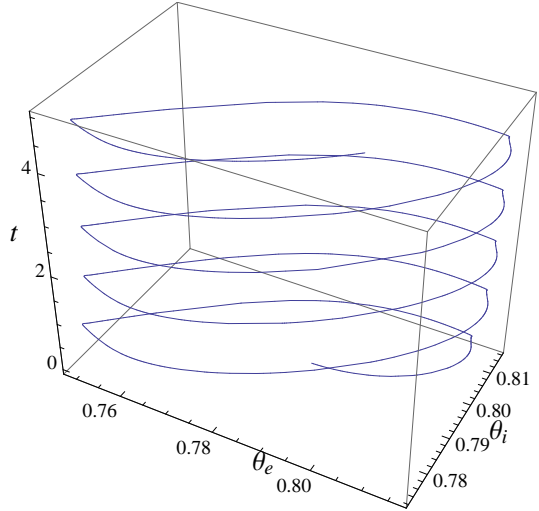}}
\caption{Numerical evolution of the two temperatures of the driven model 
corresponding to Example 1.}
\label{t-evol_2node_2}
\end{figure}

If we want to observe the evolution of $\T_\mathrm{e}$ and $\T_\mathrm{i}$
simultaneously, a three-dimensional plot is necessary. However, the plotting
of several integral curves in the same graph makes it confusing, so we choose
to plot only one, namely, the curve with initial conditions at the fixed
point. This plot is displayed in Fig.~\ref{t-evol_2node_2}.  Notice that the
amplitude of the oscillation is sufficiently small for the evolution to stay
in a small neighborhood of the sink of the averaged equations, where the
linear equations (\ref{linODE-pert}) are good approximations (compare the
amplitude with the ranges displayed in Fig.~\ref{flow1}).  The limit cycle can
be mentally visualized in this three-dimensional plot by identifying, for
example, the plane $t=5$ with the plane $t=4$.
Of course, a three-dimensional plot in cylindrical coordinates such that $t$
were the angular coordinate and the temperature planes were orthogonal to it
would be more suitable for representing the limit cycle, but that plot would
not provide a clear rendering of the line converging to the limit cycle.

Like in the one-node model \cite{NoDy}, one can derive a reasonable
approximation of the limit cycle from Eqs.~(\ref{T_1_lim}) or
(\ref{Fourier-sol}).  Since the magnitude of the eigenvalues of $J$ is not
small (they are $\{-10.74, -1.024\}$), it is more efficient to use the
integral expression (\ref{T_1_lim}), because the decrease of the integrand
with $\tau$ allows us to compute the integral to a good accuracy by
restricting it to a few periods of $f$.

It is interesting to explore what happens for other values of the parameters,
especially, for increasing magnitudes of the driving $f$ which could
compromise the convergence of the perturbation series. According to
Eq.~(\ref{f}) and on account of the proportionality between $q_\mathrm{a}$ and
$q_\mathrm{s}$ (for fixed albedo), $f$ is proportional to $q_\mathrm{s}$, that
is to say, the driving heat oscillation is proportional to the solar constant.
This constant would be larger if we considered a satellite orbiting an inner
solar system planet, for example.  We have carried out numerical integrations
with the values of $q_\mathrm{s}$, $q_\mathrm{a}$ and $q_\mathrm{p}$ (all
proportional to the solar constant) increased by a given factor. Nothing
remarkable happens with a factor as large as one hundred. However, a factor of
about one hundred ninety seems to provoke an instability. For factors larger
than 250, the periodic limit-cycle behavior seems to become quasi-periodic,
and it even seems to become chaotic for larger factors. We have not studied
the transition to chaos in any detail, because such large values of the heat
inputs are far from being realistic.

\section{Many-node model of a spacecraft}
\label{sec:2}

Let us now consider a general many-node thermal model of a spacecraft (it
could be a satellite, in particular). The energy balance equations are
\cite{anal-sat}
\begin{eqnarray}
\label{dimODE_N}
C_i \, \dot{T_i} = \dot{Q}_i(t) + 
\sum_{j=1}^N{}^{'}\left[K_{ij} (T_j - T_i) + R_{ij} (T_{j}^4 - T_{i}^4)
\right] - R_i\,{T_i}^4,
\quad i= 1,\ldots,N,
\end{eqnarray}
where $N$ is the number of nodes, and the prime in the sum symbol means that
the value $j=i$, namely, the self-coupling, is omitted.%
\footnote{In this formula, the prime is irrelevant, because one can introduce
arbitrary values of $K_{ii}$ and $R_{ii}$, for $i= 1,\ldots,N,$ and the
corresponding terms identically vanish. However, the prime is relevant in
subsequent formulas that derive from this formula.}  $\dot{Q}_i(t)$ contains
the total heat input to the $i$-node from outside of the spacecraft and from
its internal heat dissipation (if there is any).  The conductive and
radiative couplings are denoted $K$ and $R$, respectively. The couplings
between two arbitrary nodes $i$ and $j$ satisfy $K_{ij} = K_{ji}$ and $R_{ij}
= R_{ji}\,$.  The $i$-node coefficient of radiation to the environment is
given by $R_i=A_i\e_i\s$, where $A_i$ denotes the outside looking area and
$\e_i$ denotes the emissivity.  
Eqs.~(\ref{dimODE_N}) basically coincide with the ones
implemented in commercial software packages, for example, ESATAN$^\mathrm{TM}$
\cite{ESATAN}.%
\footnote{However, the emission terms $R_i\,{T_i}^4$ are
absent in ESATAN$^\mathrm{TM}$, because an extra ``environment'' node at fixed
temperature $T_0 = 3\!$ K is introduced instead ($T_0$ is the temperature of
the cosmic microwave background radiation). The effect of the extra node is
equivalent to replacing ${T_i}^4$ with ${T_i}^4-T_0^4$ in the emission terms
in Eqs.~(\ref{dimODE_N}), which has no effect if ${T_i} \gg T_0$.}

Like in the two-node case, our first step is to assume in
Eqs.~(\ref{dimODE_N}) that the $\dot{Q}_i$ are constant, by replacing the
functions $\dot{Q}_i(t)$ with their averages.  The steady state temperatures
are the roots of a system of $N$ algebraic equations of fourth degree. Systems
of algebraic equations are notably more difficult to solve than single
algebraic equations, but some general facts about them are known. For example,
the number of complex roots of a systems of $N$ fourth-degree algebraic
equations is $4^N$, generically speaking.%
\footnote{This a consequence of Bezout's theorem: ``The number of solutions of
a system of $n$ homogeneous equations in $n+1$ unknowns is either infinite or
equal to the product of the degrees, provided that their solutions are counted
with their multiplicities,'' as stated by Shafarevich~\cite{Bezout}.
The homogenization of the equations is necessary for the theorem to hold, that
is to say, the theorem actually refers to equations in projective
space. Nevertheless, we can say that the number of solutions of the
corresponding equations in the affine space $\mathbb{R}^n$ is {\em
generically} the product of the degrees.}
However, we are only interested in the roots with real and positive $T_i,\;i=
1,\ldots,N$.  The problem of finding the physical steady state for the
two-node model of Sect.~\ref{sec:1} is solved in Sects.~\ref{steady} and
\ref{stab}, where we show that, indeed, there is only one root in the positive
quadrant and it corresponds to an asymptotically stable state.  We now deal
with the general problem and prove that there is one and only one
asymptotically stable state in the positive orthant.

\subsection{Averaged equations: their stable steady state}

Let us recall the results for the two-node steady state in
Sect.~\ref{steady}. The total number of complex roots of the two algebraic
equations is indeed $4^2=16$. However, to conclude that there are 4 real roots
of which only one is in the positive quadrant, we need to reduce the two
equations to a single fourth-degree equation and apply Descartes's rule of
signs to it. Unfortunately, this is a very specific procedure tailored to
those two algebraic equations. In fact, the general two-node model given by
Eqs.~(\ref{dimODE_N}) with $N=2$ and constant $\dot{Q}_i, \;i=1,2,$ gives rise
to two fourth-degree algebraic equations that do not lend themselves to be
reduced to a single fourth-degree algebraic equation. Nonetheless, one can
reduce the two equations to a single sixteenth-degree equation; but its
analysis is inconclusive, whether we use Descartes's rule of signs or other
standard methods of determining the number of real roots of an algebraic
equation \cite{Hymers}.

Instead of attempting to find the zeros of the ODE's vector field directly by
solving algebraic equations, we can resort to an indirect method, namely, to a
{\em topological} method. For a vector field and a closed curve in the plane,
one can introduce the Poincar\'e {\em index} of the curve with respect to the
vector field, which is the number of turns that the vector makes when a point
goes along the curve and returns to its original position \cite{Andro,Arnold}.
The Poincar\'e index of a curve is a topological invariant, for it only
depends on the singularities (zeros) of the vector field enclosed by the
curve. In particular, when the curve encloses no singularities, the index
vanishes.  Therefore, a non-vanishing index proves the existence of a
singularity in the given region. In particular, we can take as
test region the positive quadrant: to convert it into a finite region, we can
bound it with a curve such that the distance of every point on it to the
origin is sufficiently large for the vector field to adopt its asymptotic
form, with only its highest degree terms. For the general two-node model, we
can then find, in particular, the expression of the vector field on the
boundary of a large square with a vertex on the origin and two sides on the
positive coordinate semi-axes, and then check that the vector always points
inwards. This shows that the index in the square is $+1$, which proves the
existence of, at least, one steady state with real and positive $T_1$ and
$T_2$.

To complete the argument, we need to determine the 
index of the possible 
steady states. 
First, let us study their stability. 
The Jacobian matrix at the point $(T_1,T_2)$ is
\begin{equation}
J = \left(
\begin{array}{cc}
[- K_{12} - 4 (R_{12}+ R_1) T_1^3]/C_1 & (K_{12} + 4 R_{12}\, T_2^3)/C_1 \\
\left(K_{12} + 4 R_{12} \,T_1^3\right)/C_2  &  
\left[- K_{12} - 4 (R_{12}+ R_2)  T_2^3\right]/C_2  
\end{array}  \right).
\end{equation}
When $R_2=0$, its eigenvalues are real and negative (provided that $T_1, T_2
>0$), as proved in Sect.~\ref{stab}.  The proof holds when $R_2 \neq 0$, so
that any fixed point in the physical region must be a sink and, specifically,
a node.  Now, we can combine this result with the available topological
information: since the Poincar\'e index of a sink is easily seen to be $+1$
\cite{Andro,Arnold} and the index is additive, there can only be one sink (in
the physical region).

The preceding analysis of the general two-node model can be extended to higher
dimensions, for the two-dimensional notion of the index of a curve with
respect to a vector field can be generalized to higher dimensions
\cite{Arnold}, giving rise to the Poincar\'e-Hopf theorem \cite{P-H}. 
Furthermore, the negativity of the eigenvalues of the Jacobian matrix also
holds in higher dimensions. However, the proof is not as simple as in the case
$N=2$. Indeed, in the general case, we need to introduce some notions of the
theory of matrices and then prove a preliminary property of the Jacobian
matrix (stated below as a {\em lemma}).

The $N$-node equations Jacobian matrix elements 
are given by:
{\setlength\arraycolsep{2pt}
\begin{eqnarray}
\label{Jij}
J_{ij} &=&  C_i^{-1} \left(K_{ij} + 4 R_{ij} T_{j}^3\right),
\quad \mathrm{if}\;i \neq j,\\
\label{Jii}
J_{ii} &=& C_i^{-1} \left[-\sum_{j=1}^N{}^{'}\left(K_{ij} + 4 R_{ij} T_{i}^3
\right) - 4 R_i\,{T_i}^3\right].
\end{eqnarray}
}%
This matrix has positive off-diagonal and negative diagonal elements (if the
temperatures are positive). This property can be expressed by saying that $-J$
is a $Z$-matrix \cite{Ber-Plem}. Furthermore, we can prove that it is also a
nonsingular $M$-matrix, namely, a $Z$-matrix such that its inverse is
non-negative. The condition that its inverse be non-negative is, in fact, just
one among a number of equivalent conditions that turn a $Z$-matrix into a
nonsingular $M$-matrix: Ref.~\cite{Ber-Plem} lists fifty different conditions!
In our case, it is convenient to apply the conditions of
{\em semipositivity} (related to the conditions of diagonal dominance,
see Ref.~\cite{Ber-Plem}).
Thus, we state:\\[1mm]
{\bf Lemma:} 
The opposite of the Jacobian matrix of Eqs.~(\ref{dimODE_N}) is 
a nonsingular $M$-matrix.\\[1mm]
{\bf Proof:} Instead of applying the semipositivity conditions to $-J$, we
apply them to its transpose, which is equivalent, since a matrix and its
transpose are simultaneously nonsingular $M$-matrices.  A matrix is {\em
semipositive} if there exists a strictly positive vector that stays strictly
positive when multiplied by the matrix.  
Let $v=(C_1,\ldots,C_N)$, so 
{\setlength\arraycolsep{2pt}
\begin{eqnarray}
\left[-J^t\cdot v\right]_{i} &=& 
-\sum_{j=1}^N{}^{'}  J_{ji}\,C_j - J_{ii} \,C_i 
= -\sum_{j=1}^N{}^{'} \left(K_{ji} + 4 R_{ji}{T_i}^3\right) +
\nonumber\\
&&+\sum_{j=1}^N{}^{'}\left(K_{ij} + 4 R_{ij} T_{i}^3
\right) + 4 R_i\,{T_i}^3  
= 4 R_i\,{T_i}^3 \geq 0.
\end{eqnarray}
}%
Therefore, the vector $-J^t\cdot v$ is strictly positive if no $R_i$ vanishes, 
but it is just positive if some of these coefficients do vanish.
Then, the matrix and the vector must fulfill additional conditions
\cite{Ber-Plem}, which are, in our case:
$$
\sum_{j=1}^i  \left(-J^t\right)_{ij} C_j > 0\,, \quad i= 1,\ldots,N.
$$
These conditions hold if we choose a node order such that $R_N \neq 0$, which
is possible, unless $R_i=0$ for all $i=1,\ldots ,N$. In conclusion, 
$-J^t$ 
and hence $-J$ are nonsingular $M$-matrices,
{\bf q.e.d.}

Once we know that $-J$ is a nonsingular $M$-matrix, we can use a
crucial property of its eigenvalues: their real parts are positive
\cite{Ber-Plem}. In fact, an $M$-matrix actually plays the r\^ole of ``a poor
man's positive definite matrix.''  In particular, the condition that the
eigenvalues of $J$ have negative real parts (in the physical region) is
sufficient to affirm the asymptotic stability of any steady state, even if the
eigenvalues have non-vanishing imaginary parts.  Therefore, any physical
steady state has to be a sink. 
Taking this into account, we can state the following theorem:\\[1mm]
{\bf Theorem:} Eqs.~(\ref{dimODE_N}) with constant $\dot{Q}_i,\; i=
1,\ldots,N,$ have in the positive orthant a unique steady state that is
asymptotically stable.\\[1mm] 
{\bf Proof:} We first prove the existence of a steady state in the positive
orthant.  Using the boundary of a large hypercube drawn from the origin in the
directions of positive $T_i$ ($i=1,\ldots,N$), it is easy to show that the
vector field defined by Eqs.~(\ref{dimODE_N}) (with constant $\dot{Q}_i$)
points inwards. Indeed, on the hypercube side $T_i=0$, the $i$-component of
the vector field is trivially positive, while on the opposite side, $T_j <
T_i$ for each $j \neq i$, making the $i$-component negative for large $T_i$.
Therefore, one deduces that the Poincar\'e-Hopf index is non-vanishing and
actually is $(-1)^N$.  Although this non-vanishing index proves the existence
of a steady state, it does not prove that it is unique.  However, we already
know that any physical steady state is a sink, and a sink has Poincar\'e-Hopf
index $(-1)^N$.  On account of the additivity of the Poincar\'e-Hopf indices,
it follows that the sink in the positive orthant is unique, {\bf q.e.d.}

Of course, we have only proved {\em local} stability. In Subsec.~\ref{stab},
we have also established the {\em global} stability of the restricted two-node
steady state by appealing to some general theorems due to the topological
restrictions of two-dimensional flows.  Thus, the proof given in
Subsec.~\ref{stab} is also valid for the general two-node model. In three or
more dimensions, there are no equivalent topological restrictions and the
question of global stability is moot.  This is natural, given that autonomous
systems of three ODE's can have very complex flows and, in particular, can
have {\em chaotic attractors}.

Since $J$ is diagonalizable with {\em real} eigenvalues in the case of the
two-node model, we would like to know if this holds for $N >2$.  Let us
consider, for example, the three-node model. It gives rise to a $3 \times 3$
Jacobian matrix and, hence, to a third-degree algebraic equation for the
eigenvalues.  If the discriminant of this equation were positive, then its
roots would be three different real numbers \cite{Tignol} and, therefore, the
Jacobian matrix would be diagonalizable (like in the case of the
two-node model).  Unfortunately, the discriminant of a third-degree algebraic
equation is a complicated fourth-degree polynomial in its
coefficients. Furthermore, the coefficients of the eigenvalue equation for a
$3 \times 3$ matrix are complicated polynomials in the matrix elements. As a
polynomial in these matrix elements, the discriminant has sixth degree and
consists of 144 monomials with positive and negative signs.  Thus, it is very
difficult to decide on the overall sign of this discriminant, despite that the
Jacobian matrix elements have definite signs.

The question of whether or not the Jacobian matrix is diagonalizable with {\em
real} eigenvalues can be considered from a different point of view.  Given the
expressions of the Jacobian matrix elements (\ref{Jij}) and (\ref{Jii}), the
matrix is simplified somewhat by multiplying it on the left by the diagonal
matrix $D= \mathrm{diag}(C_1,\ldots,C_n)$, which removes the factors
$C_i^{-1}$.  If $D \cdot J$ were symmetric, $D^{1/2} \cdot J \cdot D^{-1/2}$
would be symmetric as well, as is easily proved.  Therefore, $J$ would be
similar to a symmetric matrix and hence would be diagonalizable with real
eigenvalues.  The matrix elements of the symmetric and antisymmetric parts of
$D \cdot J$ are, respectively, $K_{ij} + 2 R_{ij} \left(T_i^3 + T_j^3 \right)$
and $2 R_{ij} \left(T_i^3 - T_j^3 \right)$ (for $i \neq j$).  If the latter
matrix element is of small magnitude with respect to the former for each pair
of nodes, then $D \cdot J$ is nearly symmetric and $J$ is likely to be
diagonalizable with real eigenvalues.  This can be deduced by considering that
the eigenvalues of a matrix and their respective eigenvectors vary
continuously with the matrix elements and that the eigenvectors of a symmetric
matrix are orthogonal. Therefore, if a variation of the elements of a
symmetric matrix is to turn two real eigenvalues into a couple of complex
conjugate eigenvalues, then it must have sufficient magnitude to change the
directions of the respective eigenvectors so much that they coincide.

It is pertinent here to comment that the procedures used in the literature for
linearizing the $N$-node ODE's~(\ref{dimODE_N}) amount to an {\em ad-hoc}
symmetrization of the matrix $D \cdot J$, either by assuming that the
steady-state node temperatures are approximately uniform
\cite{anal-sat} or, more accurately, by assuming that they fulfill 
$\left|T_i - T_j\right| \ll (T_i + T_j)/2$, for each pair of nodes
$ij$ such that $R_{ij} \neq 0$  \cite{IDR}.

\subsection{Slowest variable and convergence to the steady state}

We have proved above that the eigenvalues of the Jacobian matrix $J$ have
negative real parts and, furthermore, we have argued that they are likely
real. In this section, we focus on the eigenvalue of smallest absolute value,
corresponding to the slowest thermal mode, and we prove that, indeed, it is
real.  The slowest mode is especially important because it eventually
determines the dynamics and the convergence to the steady state.  For the
proof, we appeal to Perron's theorem for positive matrices: a positive
matrix has a unique real and positive eigenvalue with a strictly positive
eigenvector and, furthermore, that eigenvalue has maximal modulus among all
the eigenvalues \cite{Ber-Plem}. Since the inverse of $-J$ is non-negative, it
is a good candidate for Perron's theorem: if it is actually strictly positive,
its maximal modulus eigenvalue corresponds to the minimal modulus eigenvalue
of $J$.

However, it is not easy to find whether or not the inverse of $-J$ is strictly
positive.  We can apply instead the following theorem: an {\em irreducible}
$Z$-matrix is a nonsingular $M$-matrix if and only if its inverse is {\em
strictly} positive \cite{Ber-Plem}.  A matrix is said to be reducible if it
can be put in a block upper-triangular form by a simultaneous permutation of
its files and columns.%
\footnote{The reducibility of a matrix can be expressed
in terms of its graph, namely, the directed graph on $N$ nodes in which there
is a directed edge leading from the node $i$ to the node $j$ if and only if
the corresponding matrix element is non-vanishing.  A graph is called {\em
strongly connected} if for each pair of nodes there is a sequence of directed
edges leading from the first node to the second node.  A matrix is irreducible
if and only if its graph is strongly connected.  Of course, the graph of the
Jacobian matrix is the one defined by the node model and the corresponding
heat exchanges.}
Therefore, we only need to show that $J$ is irreducible. It would certainly
be so if $K_{ij}$ or $R_{ij}$ did not vanish for any pair of indices
(except when $i=j$); but some (or many) of the conductive and radiative
couplings are expected to vanish simultaneously.  Let us assume that we can
relabel the nodes and hence permute simultaneously the files and columns of
$J$ to make it block upper-triangular. Given that an element $J_{ij}$ only
vanishes if both $K_{ij}$ and $R_{ij}$ vanish, according to Eq.~(\ref{Jij}),
both elements $J_{ij}$ and $J_{ji}$ vanish or do not vanish simultaneously.
In consequence, an order of nodes that makes $J$ block upper-triangular also
makes it block diagonal.  A matrix that can be put in a block diagonal form by
a simultaneous permutation of its files and columns is said to be {\em
completely reducible}. However, $J$ is completely reducible if and only if the
node model splits into two disconnected node models, which corresponds to
modelling two separated spacecrafts.

In conclusion, the matrix $J$ is irreducible for a single spacecraft model and
we can apply Perron's theorem to $(-J)^{-1}$: the largest eigenvalue of the
latter corresponds to the negative eigenvalue of $J$ of smallest magnitude
and, therefore, to the slowest variable. Furthermore, the corresponding
eigenvector is positive and, actually, the only positive eigenvector.
Therefore, 
the sink is eventually approached either from the
zone corresponding to simultaneous temperature increments or from the zone
corresponding to simultaneous temperature decrements, 
as already observed for the two-node models in Sect.~\ref{num}.  
In other words, 
the slow mode corresponds to a simultaneous increase or decrease of
the (non-uniform) temperature throughout the spacecraft whereas faster modes
correspond to temperature increases in one or more parts of the spacecraft
that are accompanied by decreases in other parts.

\subsection{Driven many-node model}

If we assume that the heat inputs $\dot{Q_i}(t)$ are periodic, the
non-autonomous equations~(\ref{dimODE_N}) is a generalization of the driven
two-node model studied in Sects.~\ref{forced} and \ref{pertur}.  Thus, we can
use the equations in Sect.~\ref{pertur}, if we replace the $\T_i$ with $T_i$
and use the driving term
$$
F_i(t) = \frac{\dot{Q_i}(t) - \langle{\dot{Q_i}}\rangle}{C_i}\,, 
\quad i=1, \ldots, N,
$$
where $\langle{\dot{Q_i}}\rangle$ is the mean value of $\dot{Q_i}(t)$ over the
period of oscillation.  The first order perturbative correction $T_{(1)j}(t)$
to the solution of the averaged equations is the solution of the system of
linear ODE's~(\ref{linODE-pert}), where $\T_{(1)j}$ must be substituted by
$T_{(1)j}$. The subsequent steps are independent of the dimension. Therefore,
the integral expression (\ref{T_1_lim}) of the limit cycle holds (after
replacing $\T$ with $T$).

If we use the basis of the eigenvectors of $J$ (assuming that $J$ is
diagonalizable), the limit cycle ${T}_{(1)}^\infty(t)$ becomes a sum of
contributions, each one corresponding to an eigenvector.  In particular, if
$f$ is the component of $F$ along some eigenvector with an eigenvalue $\l$
that is large compared with the heat-input frequency, the corresponding
contribution to ${T}_{(1)}^\infty(t)$ is approximately equal to $f(t)/\l$.
In a many-node model such that the magnitudes of the $J$-eigenvalues span a
long range, surely only a few of the slowest modes make significant
contributions to ${T}_{(1)}^\infty(t)$.  This observation suggests an
effective method of computing a numerical approximation of the limit cycle:
one is to begin with the contributions of the slowest modes and keep adding
modes until the addend becomes negligible with respect to the partial sum.

As regards the full perturbation series $\sum_i {T}_{(i)}^\infty(t)$, one
could be more concerned about its convergence now than in the two-node case,
since we cannot affirm that the steady state of the averaged system is
globally stable.  However, exploratory numerical work carried out for models
with few nodes for various values of the parameters suggests that the steady
state of the averaged system is globally stable and that the perturbation
series has good convergence properties.  Of course, these numerical results
have heuristic value only.

\section{Summary and discussion}
\label{sec:4}

This work is devoted to the study of the nonlinear ODE's employed in
spacecraft thermal control, with emphasis on analytic methods.  We have first
studied a satellite model consisting of two nodes, namely, the satellite's
exterior and interior parts.  Comparing the results obtained for this model
with the results for the one-node model in Ref.~\cite{NoDy}, we see that the
addition of one node does not give rise to any essentially new features, as
long as the heat inputs vary moderately.  In fact, with constant heat input
and when the thermal coupling between the two nodes is sufficiently strong, as
in Example 1 in Subsect.~\ref{Example2}, the two-node model naturally evolves
towards the one-node model. This evolution pattern is ascertained by two
results: (i) the existence of a globally stable steady state in which the
temperatures of the two nodes become almost equal; and (ii) the appearance of
a fast dynamical variable, such that the two-node model dynamics converges to
a one-node model dynamics before reaching the steady state.

When the coupling between the two nodes is weak, as in Example 2 in
Subsect.~\ref{Example2}, there is also a physical steady state and a fast
variable, but the heat balance is reached in a more complicated way.  The
ratio of the inner to the outer node temperatures in the steady state is still
close to one, e.g., $T_\mathrm{i}/T_\mathrm{e} = 1.116$ in Example 2, but the
ratio between the increments of the respective temperatures, given by the
positive eigenvector, can now be considerable: it is $\D T_\mathrm{i}/\D
T_\mathrm{e} = 3.103$ in Example 2. Therefore, the transition to the slow
dynamics does not imply that the node temperatures approach each other over
the typical time of the fast variable, although they do so over the typical
time of the slow variable.

A periodic heat input produces nonlinear driven oscillations of the
temperatures, namely, a limit cycle. This can be proved by taking the
oscillating heat input as a perturbation and expanding the nonlinear system in
series, thus converting it into a set of linear systems.  At the first order,
the linear equation valid on long times is just the equation of driven
overdamped linear oscillations and it has the standard solution.  We find that
the perturbative series is convergent in a range of amplitudes of the driving
heat input that is sufficient for the applications.  Moreover, the
perturbative series allows one to calculate, order by order, the limit cycle
and the convergence of the temperatures to it. We have found numerical
evidence of more complex behaviors, probably including quasiperiodicity and
chaos, but these behaviors take place for unrealistic amplitudes of the
driving heat input.

Most of the above conclusions can be extended to the general $N$-node thermal
model of a spacecraft, by employing topological methods and the theory of
non-negative matrices.  For constant heat input, the general $N$-node model
also has an attractive steady state in the physical region and the dynamics
converges to a slow variable corresponding to a simultaneous increase or
decrease of the (non-uniform) temperature throughout the spacecraft.  However,
we are unable to prove the global stability of the steady state.  In this
regard, we must notice that most proofs in the theory of differential
equations have local nature and, in fact, global properties can only be proved
with powerful mathematical tools. One such tool is the topological index, but
it only provides partial information. Another powerful tool is the existence
of a Lyapunov function \cite{Gu-Ho,Drazin}: a global strict Lyapunov function
allows one to determine the global stability of a sink. Unfortunately, there
are no general methods for finding suitable Lyapunov functions. For ODE's with
physical origin, sometimes it is possible to find a physically motivated
Lyapunov function, for example, an energy, entropy, etc. Thus, there could be
a physically motivated Lyapunov function for the $N$-node model ODE system,
but it is certainly hard to find.

When driven by a periodic heat input, the $N$-node model also becomes a
nonlinear oscillator.  But one could be more concerned about the range of
convergence of the perturbative series for this model than about the series
for the two-node model.  At any rate, when the dynamics is confined to a
neighborhood of the steady state, no high order terms of the perturbative
series are needed.  The amplitude of the two-node model limit cycle is indeed
small and the linear first-order equation suffices; but its steady state is,
of course, globally stable anyhow.  Regarding $N$-node models, we believe that
the relevant values of the parameters do not produce large amplitudes either
and that the linear first-order equations may suffice.  Notice, however, that
those equations do not coincide with previous types of linearization, in which
the coefficient matrix of the ODE system is forced to be related to a
symmetric matrix by defining radiative conductances that depend on somewhat
arbitrary temperatures (as in Refs.~\cite{anal-sat,anal-sat_1,IDR}).  In
contrast, the linear equations are not equivalent to the equations of a model
with only conductive thermal couplings and the actual Jacobian matrix of the
ODE system is not related to any symmetric matrix.

Regarding the practical use of $N$-node thermal models, we must notice that
the diagonalization of the Jacobian matrix at the steady state usually yields
(negative) eigenvalues of very different magnitude, like in our two examples
of a two-node model.  The fast variables indicate the regions of the
spacecraft where the heat inputs and outputs quickly balance, establishing a
dynamical balance long before the steady state is reached.  The last and
longest stage of this dynamical balance corresponds to the slowest variable,
with definite ratios of the temperature increments.  We believe that it is
interesting to distinguish the response of the different modes in regard to
spacecraft thermal control.  Standard software packages for spacecraft thermal
analysis do not regard this aspect of the solution of the equations.

Besides, our results have some bearing on an interesting problem of
lumped-parameter models, namely, the problem of node condensation: a model is
normally constructed heuristically and it can be useful to reduce its number
of nodes while preserving the required performance of its design.  In a
spacecraft thermal model, if a couple of nodes $ij$ are such that their steady
temperatures fulfill $T_i \simeq T_j$ and $\D T_i
\simeq \D T_j$, it may be advisable to replace them by a single node, because
that will not result in a loss of accuracy in the description of the
temperature distribution. According to our results, this type of condensation
can be achieved at no great computational cost: one only needs to compute the
steady temperatures and then the positive eigenvector of the corresponding
Jacobian matrix.

The determination of a spacecraft's thermal modes and their use for node
condensation are beyond the scope of current thermal software
packages. Therefore, the analytic approach proposed here can be a useful
complement to the analysis that is normally carried out with those software
packages.

 \subsection*{Acknowledgments}
I thank Angel~Sanz-Andr\'es for conversations.

\end{document}